\documentclass[journal]{IEEEtran}

\ifCLASSINFOpdf
\else
   \usepackage[dvips]{graphicx}
\fi
\usepackage{url}
\usepackage{graphicx}
\usepackage[justification=centering]{caption}
\usepackage{multirow}
\usepackage{amsmath}
\usepackage{upgreek}

\begin{document}

\title{Watermarking for Neural Radiation Fields by Invertible Neural Network}

\author{Wenquan Sun, Jia Liu, Weina Dong, Lifeng Chen and Ke Niu
\thanks{Manuscript received 2 December 2023; accepted X December 2023. Date of
publication X X 2023. This work was supported in part by the General Program of the National Natural Science Foundation of China under Grant 62272478.The associate editor coordinating the review of this manuscript and approving it for publication was XXX (Corresponding author: Jia Liu.)}
\thanks{The authors are with the College of Cryptography Engineering, Engineering University of PAP, Xi’an Shanxi 710086, China; Key Laboratory of Network and Information Security of PAP (Engineering University of PAP), Xi’an Shanxi 710086, China. (e-mail: 1240464419@qq.com;liujia1022@gmail.com;703792110@qq.com;
3011745933@qq.com;niuke@163.com).}
}
\markboth{Journal of \LaTeX\ Class Files, Vol. 14, No. 2, December 2023}
{Shell \MakeLowercase{\textit{et al.}}: Bare Demo of IEEEtran.cls for IEEE Journals}
\maketitle

\begin{abstract}
To protect the copyright of the 3D scene represented by the neural radiation field, the embedding and extraction of the neural radiation field watermark are considered as a pair of inverse problems of image transformations. A scheme for protecting the copyright of the neural radiation field is proposed using invertible neural network watermarking, which utilizes watermarking techniques for 2D images to achieve the protection of the 3D scene. The scheme embeds the watermark in the training image of the neural radiation field through the forward process in the invertible network and extracts the watermark from the image rendered by the neural radiation field using the inverse process to realize the copyright protection of both the neural radiation field and the 3D scene. Since the rendering process of the neural radiation field can cause the loss of watermark information, the scheme incorporates an image quality enhancement module, which utilizes a neural network to recover the rendered image and then extracts the watermark. The scheme embeds a watermark in each training image to train the neural radiation field and enables the extraction of watermark information from multiple viewpoints. Simulation experimental results demonstrate the effectiveness of the method.
\end{abstract}

\begin{IEEEkeywords}
Neural Radiance Fields, Copyright Protection, 3D Scene, Watermarking, Invertible Neural Networks.
\end{IEEEkeywords}

\IEEEpeerreviewmaketitle

\section{Introduction}

\IEEEPARstart{I}{mplicit} Neural Representation (INR), also known as a coordinate-based representation, is a method for parameterizing various signals. While traditional signal representations are usually discrete, implicit neural representations parameterize a signal as a continuous function. Currently, the most typical application of INR is Neural Radiance Fields (NeRF)\cite{mildenhall_nerf_2020}. NeRF is a deep learning model for 3D implicit spatial modeling that uses neural networks to implicitly represent the color and density functions of each point in a 3D scene. Current research in NeRF is dedicated to working on higher quality 3D content representation \cite{wang_neus_2021,zhang_nerfactor_2021,barron_mip-nerf_2021,tancik_block-nerf_2022}, faster rendering \cite{schwarz_voxgraf_2022,yu_plenoxels_2021,muller_instant_2022,wang_fourier_2022,sun_direct_2021}, and sparse view reconstruction \cite{xu_signal_2022,chen_geoaug_2022,chen_mvsnerf_2021,yu_pixelnerf_2020,zhang_ray_2022,niemeyer_regnerf_2021}. As NeRF continues to progress in 3D representation, the issue of copyright protection for 3D models of neural radiance fields oriented towards implicit representation has become a pressing topic.

Traditional representations of 3D models can be categorized into point cloud models, mesh models, and surface models. Watermarking techniques for traditional 3D models can be mainly classified into two categories: 3D mesh model-based watermarking algorithms \cite{qin_effective_2015,liao_reversible_2015,uccheddu_wavelet-based_2004,praun_robust_1999,ohbuchi_frequencydomain_2002,hou_blind_2017,son_perceptual_2017,hamidi_blind_2019} and 3D point cloud model-based watermarking algorithms. For 3D mesh model-based watermarking algorithms, a multi-resolution framework is typically used to perform wavelet decomposition or Fourier transform on the target triangular or polygonal mesh. The watermark embedding is achieved by modifying the topological or geometric features of the mesh model or establishing a correlation function between the mesh vertices. On the other hand, the 3D point cloud model watermarking algorithm \cite{liu_novel_2019} first establishes a synchronization relationship between point clouds. Then, the model is divided into spherical rings based on the radial radius, and the watermark is repeatedly inserted into the vertices of each sphere ring to realize the watermark embedding.

However, NeRF's 3D model representation differs from traditional 3D models in that NeRF does not use geometric structures in the traditional sense, but instead learns and generates realistic renderings directly through neural networks. It is essentially a neural network that performs an implicit representation of the 3D scene. Therefore, traditional 3D model watermarking algorithms cannot be applied to watermark neural radiation fields. Copyright protection for neural networks, i.e., neural network watermarking, has become an important research direction in the security field. There are four main types of watermarking for neural networks: white-box watermarking, black-box watermarking, no-box watermarking, and vulnerable neural network watermarking. In the white-box watermarking scheme \cite{uchida_embedding_2017}, the verifier can access the interior of the network and retrieve information such as weights when verifying the copyright of the network. The black-box watermarking scheme \cite{adi_turning_2018} is suitable for cases where the verifier cannot access the interior of the network but can only interact with the network through a remote API interface. Boxless watermarking \cite{wu_watermarking_2021} is primarily used for copyright authentication in generative networks, where the approach involves training the network in such a way that the generated image contains watermark information, which the verifier can then directly verify for copyright. Vulnerable watermarking \cite{guan_reversible_2021} differs from the above three methods in that it detects whether the functionality of the network has been maliciously tampered with, such as through the injection of a backdoor, by examining the corruption of the watermark.

Traditional neural network watermarking typically considers the neural network as a tool for data processing, thus aiming to protect the tool itself. However, in the context of implicit expression, the neural network itself becomes the data. Therefore, this paper proposes that future protection of multimedia data (such as images, videos, audio, etc.) can be approached in two ways, as depicted in Figure 1. The first approach involves directly applying traditional watermarking techniques to safeguard multimedia data. The second approach involves first transforming the multimedia data into neural network data using neural implicit expression, and then utilizing neural network watermarking techniques to protect the multimedia data represented by the implicitly expressed network.
\begin{figure}
\centerline{\includegraphics[width=\columnwidth]{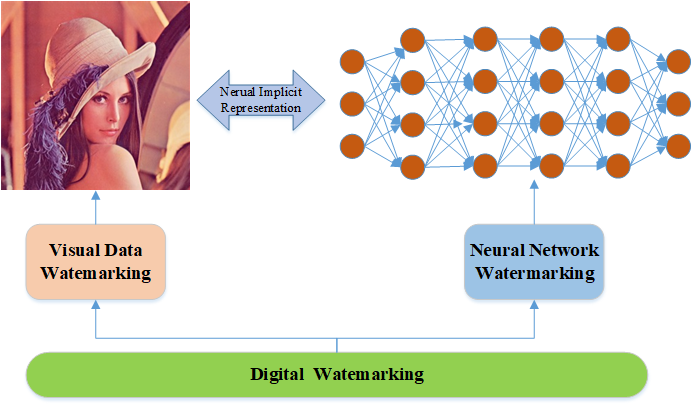}}
\caption{Two ways of protecting data.}
\end{figure}

StegaNeRF \cite{li_steganerf_2022} was the first to establish a connection between neural radiation field and message hiding by training NeRF twice to extract messages from rendered images. In contrast to StegaNeRF, this paper presents a novel approach to safeguarding the neural radiation field by employing invertible neural network watermarking. This technique does not modify the NeRF network but ensures the protection of the NeRF model by leveraging traditional image watermarking techniques. The proposed scheme starts by utilizing the forward network watermarking algorithm of invertible neural networks to embed watermark information into each image separately within the training set used for NeRF training. Following this, 3D modeling is conducted using the NeRF model. To counteract the influence of NeRF rendering, the verifier can render the NeRF from any viewpoint and subsequently recover the rendered image using a trained image quality enhancement network. Finally, the verifier can extract the embedded watermark information using the reverse process of the invertible network, namely the extraction network. In the case of a black-box scenario, where the 3D model is suspected to be utilized by unauthorized parties, the verifier can extract watermark information from multiple perspectives to verify the network copyright.

\section{The algorithms in this paper}
\begin{figure*}[ht]
\centerline{\includegraphics[height=5cm,width=7.5in]{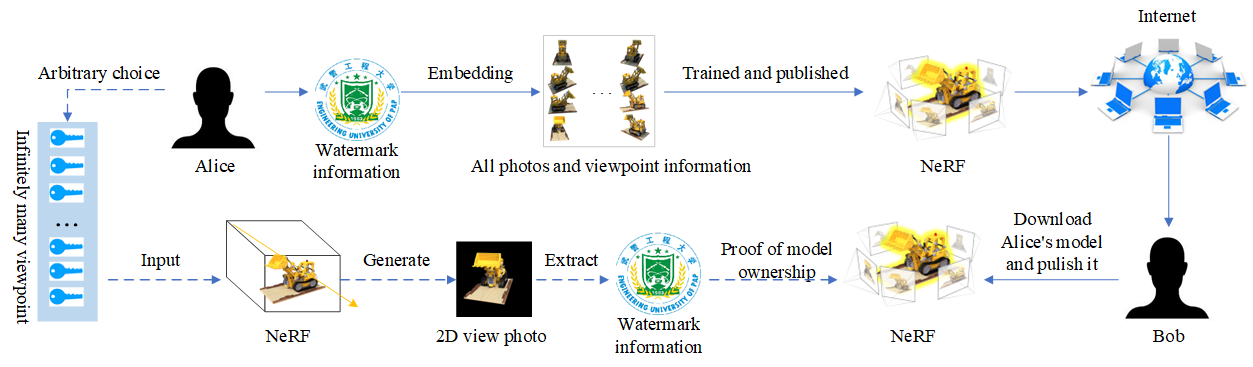}}
\caption{Application scenario flowchart}
\end{figure*}
\subsection{Application Scenarios}
The usage scenarios of the algorithm described in this paper are as follows:

Alice has acquired some pictures of a 3D scene by taking photographs, etc;

Alice embeds watermarks in the images and renders 3D scenes by training NeRF models;

Alice shares NeRF models and 3D scenes online for others to enjoy;

Bob acquired the NeRF model without Alice's permission and posted it on the web under his name;

Alice sees the NeRF model posted by Bob and uses his model to render a 2D image for watermark extraction, thereby verifying that Alice is the copyright holder of the NeRF model;

Bob is infringing on copyright and needs to withdraw the release;

\subsection{General framework of the algorithm}
In this paper, we propose a new scheme using invertible neural network 2D watermarking algorithm to achieve the protection of neural radiation field and 3D scene. As shown in Fig. 3 the algorithm framework contains frequency domain transform module, invertible module, neural radiation field and image quality enhancement module composition. Embedding and extraction in invertible neural network watermarking is a pair of inverse processes
\begin{equation}\label{}
{I_W} = H\left( {I,{M_W}} \right)   
\end{equation}

\begin{equation} \label{2)} 
\left(I^{''} ,R_{W} \right)=H^{-1} \left(QEM\left(NeRF\left(I_{W} \right)\right)\right) 
\end{equation} 
(1)-(2) where: \textit{H(-)} represents the forward embedding watermarking process, and $H^{-1}$(-) denotes the reverse extraction watermarking process. In the forward embedding watermarking process, the training image \textit{I} and the watermark information $M_{W}$ serve as inputs. These inputs are initially decomposed into high-frequency and low-frequency wavelet subbands through discrete wavelet transform (DWT) within a sequence of invertible blocks. After the final invertible block input, the inverse wavelet transform is performed using IWT to generate the watermarked image IW and the loss information \textit{r}. All training images for NeRF should undergo these operations to ensure that the watermarking information can be extracted from any angle in the training set. The resulting watermark image $I_{W}$ is then utilized to train the NeRF model, specifying the camera position, orientation, and field of view parameters for rendering. The rendered image \textit{I'} is generated using ray-voxel intersection sampling and color blending operations. In the reverse extraction watermarking process, the rendered image \textit{I'} initially undergoes the image quality increase module (QEM) to mitigate distortion effects caused by the NeRF rendering process. Subsequently, similar to the embedding process, the auxiliary variable \textit{Z} and the quality-enhanced rendered image \textit{I'} are subjected to a frequency domain transform and a series of invertible blocks to generate the recovery watermark  and the recovery image $I_{r}$.

\subsection{Network structure}

\begin{figure*}[tb]
\centerline{\includegraphics[height=4cm,width=7.5in]{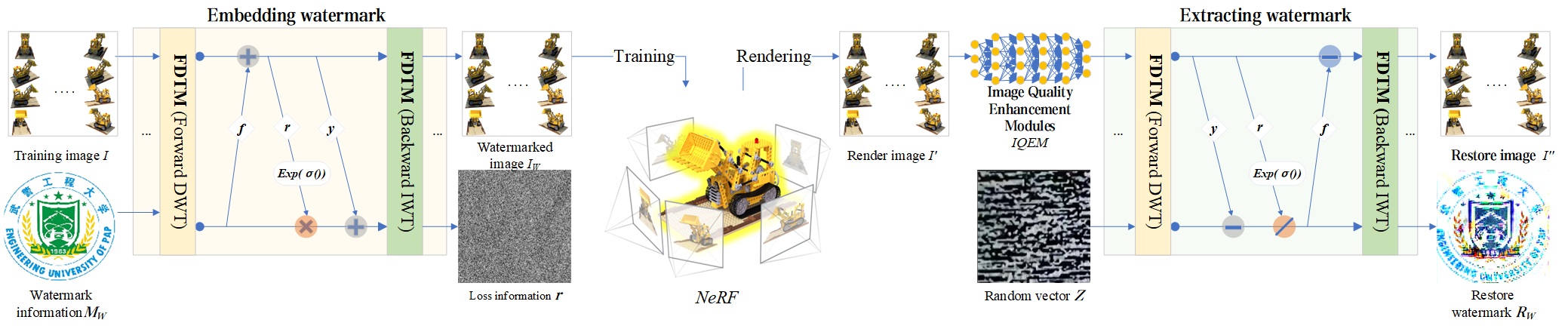}}
\caption{Basic algorithmic framework.}
\end{figure*}

\subsubsection{Frequency domain transform module}

Watermarked images embedded in the pixel domain are prone to texture replication artifacts and color distortion \cite{fridrich_detecting_2001,weng_high-capacity_2019}. The frequency domain and high-frequency domain are better suited for watermark embedding compared to the pixel domain. This paper utilizes the frequency domain transform module (FDTM) to partition the image into low-frequency and high-frequency wavelet subbands before the invertible transform. The high-frequency subbands contain the image details, while the low-frequency subbands encompass the overall image features. This division allows the network to effectively integrate the watermark information into the cover image. The wavelet transform, when compared to direct operations in the original image domain, offers improved visual fidelity and embedding efficiency as it operates on only a few subbands. Consequently, this approach minimizes the impact on the entire image and proves generally difficult to detect. Moreover, the favorable reconstruction properties of wavelets \cite{mallat_theory_1989} contribute to reducing information loss and enhancing watermark embedding capabilities. Before entering the invertible block, the image undergoes the FDTM, and following the discrete wavelet transform (DWT), the feature map's size (B, C, H, W) is transformed into (B, 4C, H/2, W/2), where B represents the batch size, H denotes the height, W indicates the width, and C represents the number of channels. The DWT reduces computational costs, thereby accelerating the training process. Subsequently, after the last invertible block, the feature map (B, 4C, H/2, W/2) is fed into the FDTM for inverse wavelet transform (IWT), resulting in the generation of the watermarked image $I_{W}$ by restoring the feature map size to (B, C, H, W).
\subsubsection{Invertible blocks}

As shown in Fig. 3, the hiding process and the recovery process have the same sub-blocks and share the same network parameters, except that the information flow is in the opposite direction. The network structure in this paper has 8 invertible blocks with the same structure, constructed as follows: For the $L^{th}$ hidden block in the forward process, the inputs are $I_{l}$ and $M_W^{l}$, and the outputs are $I^{l + 1}$ and $M_W^{l + 1}$.
\begin{equation} \label{3)} 
I_{}^{l + 1} = I_{}^l + f\left( {M_W^l} \right)
\end{equation}
\begin{equation} \label{4)} 
M_W^{l + 1} = M_W^l \otimes \exp \left( {\sigma \left( {r\left( {I_{}^{l + 1}} \right)} \right)} \right) + y\left( {I_{}^{l + 1}} \right)
\end{equation}

Equations(3)-(4): The activation function \textit{$\sigma$} is utilized, specifically the LeakyReLU. The densely connected networks, denoted as\textit{ f(-)}, \textit{r(-)}, and \textit{y(-)}, are applied to invertible block, the outputs of the final invertible block are \textit{M${}^{k}$${}_{W}$} and \textit{I${}^{k}$}. These outputs are further transformed using the inverse wavelet transform (IWT) to obtain the dense-containing image, \textit{I${}_{W}$}, and the loss information, \textit{r}. The \textit{L${}^{th}$} display block in the reverse recovery process takes inputs \textit{I${}^{'l+1}$} and \textit{Z${}^{l+1}$} and produces outputs \textit{I${}^{'}$${}_{l}$} and \textit{Z${}^{l}$}. The equations(5)-(6) illustrate this block as follows.
\begin{equation}\label{5)}
{Z^l} = \left( {{Z^{l + 1}} - y\left( {I_{r}^{l + 1}} \right)} \right) \otimes \exp \left( { - \sigma \left( {r\left( {I_{r}^{l + 1}} \right)} \right)}\right)
\end{equation}
\begin{equation}\label{6)}
I _{r}^l = I _{r}^{l + 1} - f\left( {{Z^l}} \right)  
\end{equation}

\subsubsection{Neural Radiation Field} 

Neural Radiation Field is a neural network model designed for generating 3D scenes. The network structure comprises multiple layers of perceptrons, which are employed to encode the scene's surface. Figure 5 illustrates this network structure.
\begin{figure}
\centerline{\includegraphics[width=\columnwidth]{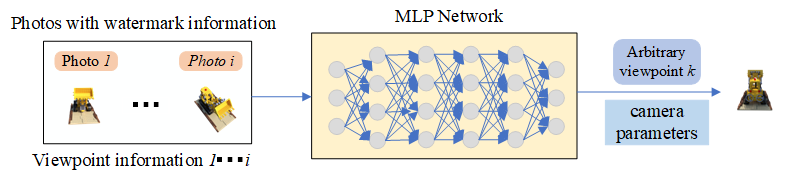}}
\caption{NeRF workflow diagram.}
\end{figure}
In the neural radiation field model, each pixel position of the input image can be represented as a 3D coordinate point in the scene, allowing for precise object location and rendering within the scene. In NeRF, the input spatial point is defined by a 3D coordinate position, \textit{x}=(\textit{x}, \textit{y}, \textit{z}), and a direction, \textit{d}=(\textit{$\theta$}, \textit{$\phi$}), while the output spatial point is characterized by a color, \textit{c}=(\textit{r}, \textit{g}, \textit{b}), and density, \textit{$\sigma$}, at the corresponding voxel position.
\begin{equation}
{F_\Theta }:(x(x,y,z),d(\theta ,\Phi )) \to (c(r,g,b),\sigma )    
\end{equation}
NeRF takes in a finite sequence of discrete images and camera parameters associated with specific viewpoints to generate a continuous static 3D scene. Moreover, it can render the scene from infinite perspectives, resulting in new viewpoint images. Body rendering, on the other hand, is a 3D-to-2D modeling process that leverages the pixel values \textit{c} and the body density \textit{$\sigma$} of 3D points obtained through 3D reconstruction. The final pixel values of the 2D image are derived by the weighted superposition of pixel point samples along a ray in the direction of observation. This process is illustrated in equation (8).
\begin{equation}
    \begin{array}{l} {T(t)=\exp (-\int _{t_{n} }^{t_{f} }\sigma (r(s)) ds),} \\ {C(r)=\int _{t_{n} }^{t_{f} }T(t)\sigma (r(t))c(r(t),d)dt } \end{array}
\end{equation}
Equation (8) introduces the ray, denoted as \textit{r(t)}, which is defined as \textit{r}(\textit{t}) = \textit{o} + \textit{t${}_{d}$}. \textit{o} represents the position of the camera's optical center, and \textit{d} represents the direction of the viewing angle. Furthermore, \textit{T}(\textit{t}) indicates the cumulative transmittance of the ray as it travels from the proximal point tn to the distal boundary \textit{t${}_{f}$}.

Building upon this characteristic of NeRF, this paper proposes a method for extracting watermarks from any angle in the training set by randomly selecting camera parameters. This approach aims to provide copyright protection for NeRF.

\subsubsection{Image Quality Enhancement Module} 

Before the reverse process to extract the watermark, to eliminate the impact of the distortion changes brought about by the NeRF rendering process, this paper sets up an image quality enhancement module (IQEM), using the residual convolutional codec network on the left side of the left 6 is a convolutional encoder, extracting the distorted image \textit{I${}^{'}$ }different levels of feature information. Then the features are input into the right side of the inverse convolutional decoder while inputting the residuals passed from the previous layer, and the final result is superimposed on the original image, which completes the image restoration. By adding IQEM to the watermark extraction process, the rendered image \textit{I${}^{'}$} is preprocessed before it enters the invertible neural network to ensure that the inputs passed backward are sufficiently similar to the watermarked image \textit{I${}_{W}$} so that the invertible neural network can extract the watermark information more completely.

\begin{figure}
\centerline{\includegraphics[width=\columnwidth]{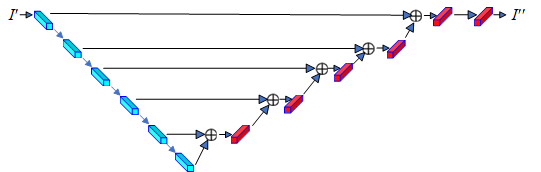}}
\caption{Image quality enhancement module architecture.}
\end{figure}

\subsection{Loss Functions}

The loss associated with network model training proposed in this paper consists of four main components:

\paragraph{ Embedding loss L${}_{E}$${}_{mb}$}

The purpose of the embedding loss is to ensure that the generated watermarked image \textit{I${}_{W}$} is indistinguishable from the training image \textit{I}. The embedding loss is used in the following steps:

\begin{equation} \label{9)} 
L_{Emb} \left(\theta \right)=\sum _{n=1}^{N}\ell _{Emb}  \left(I_{W}^{\left(n\right)} ,I_{}^{\left(n\right)} \right) 
\end{equation} 

In Eq.(9), \textit{N }represents the number of training samples, and \textit{$\ell$${}_{Emb}$} calculates the difference between the watermarked image \textit{I${}_{W}$} and the training image \textit{I}. In this paper, we use the \textit{L${}_{2}$} paradigm.

\paragraph{ Low-frequency wavelet loss L${}_{low-f}$}

Literature\cite{baluja_hiding_2017} verifies that watermark information embedded in high-frequency components is less detectable than watermark information embedded in low-frequency components. To ensure higher visual fidelity and minimize the impact on the image as a whole due to the embedding of the watermarking information, so that the watermarking information is embedded in the high-frequency region of the image as much as possible, this paper employs a loss constraint on the low-frequency subbands of the training image\textit{ I} and the watermarked image \textit{I${}_{W}$}.

\begin{equation} \label{10)} 
L_{low-f} \left(\theta \right)=\sum _{n=1}^{N}\ell _{f}  \left(H\left(I_{}^{\left(n\right)} \right)_{ll} ,H\left(I_{W}^{\left(n\right)} \right)_{ll} \right) 
\end{equation} 

In Eq.(10), \textit{N} represents the number of training samples, \textit{$\ell$${}_{f}$ }calculates the low-frequency difference between the training image \textit{I} and the watermarked image\textit{ I${}_{W}$}, and \textit{H(-)${}_{ll}$}${}_{\ }$represents the low-frequency subband operation of the extracted image.

\paragraph{ Extraction loss L${}_{Ext}$}

To ensure the consistency between the extracted watermark information \textit{R${}_{W}$} and the embedded watermark information \textit{M${}_{W}$}. The difference between the recovered watermark \textit{R${}_{W}$} and the embedded watermark information \textit{M${}_{W}$} is minimized to improve the watermark extraction accuracy of the model.
\begin{equation} \label{11)} 
L_{Ext} \left(\theta \right)=\sum _{n=1}^{N}E_{z\sim p\left(z\right)}  \left[\ell _{Ext} \left(R_{W}^{\left(n\right)} ,M_{W}^{\left(n\right)} \right)\right] 
\end{equation} 
In Eq.(11), \textit{N} represents the number of training samples, and \textit{$\ell$${}_{Ex}$${}_{t}$} computes the difference between the watermark information \textit{M${}_{W}$} and the recovered watermark \textit{R${}_{W}$}. The process of sampling the random vector \textit{z} is random.

The total loss function of the invertible neural network is a weighted sum of the embedding loss, the low-frequency wavelet loss, and the extraction loss:
\begin{equation} \label{12)} 
L_{total} \left(\theta \right)=\lambda _{1} L_{Emb} +\lambda _{2} L_{low-f} +\lambda _{3} L_{Ext}  
\end{equation} 
In the training process, \textit{$\lambda$${}_{2}$} is first set to 0, i.e., the network model is directly pre-trained without considering the effect of  \textit{L${}_{low-f}$} on the network, so that the network model first obtains the basic embedding-extraction ability. Then the \textit{L${}_{low-f}$} constraints are gradually added to further optimize the network model to embed the watermark information in the high-frequency region of the training image, to minimize the impact of the embedding of the watermark information on the image as a whole.

\paragraph{ Loss of image quality enhancement module MSE}

To ensure that the embedding of the watermark does not damage the original 2D image content, the image quality enhancement module in this paper is independent of the training of the invertible neural network, and the loss of the image quality increase module is constrained by the MSE, which is designed to ensure that the image \textit{I'} rendered by the NeRF can be restored to the watermarked image \textit{I${}_{W}$} generated by the invertible neural network to resist image watermark corruption and loss caused by the rendering process.
\begin{equation}\label{13)}
MSE = \frac{1}{n}\sum\limits_{i = 1}^n {\left( {{I_{i}^{'}},{I_{Wi}}} \right)}    
\end{equation}
Eq.(13) where \textit{I'${}_{i}$} is the \textit{i${}^{th}$} rendered image and \textit{I${}_{Wi}$} is the \textit{i${}^{th}$} watermarked image.

\section{Experimental results and analysis}
In this study, the network model employed the Pytorch platform with Cuda version 11.6 and Nvidia GeForce RTX2070 GPU. The NeRFSynthetic datasets of Lego, Hotdog, and Chair were used to train the Nerf model. To ensure diversity, high resolution, and authenticity, the DIV2K dataset was used for training the invertible neural network structure, which was modified from HiNet \cite{jing_hinet_2021}. Specifically, the DIV2K training dataset, consisting of 800 images with a resolution of 1024$\times$1024, was used for training, while the validation dataset (100 images, resolution of 1024$\times$1024) was used for validating the network model. To test the effectiveness of the network model, the DIV2K test dataset (100 images, resolution of 1024$\times$1024) was used. The Adam optimizer was used with $\uplambda$${}_{1}$=5, $\uplambda$${}_{2}$=0.5, $\uplambda$${}_{3}$=1, a learning rate of 1$\times$10${}^{-4.5}$, and a batch size of 2 for the network model training. The entire network model consisted of 8 invertible blocks, with each block containing three DenseNet blocks including 7 layers of convolutional blocks as \textit{f(-)}, \textit{r(-),} and \textit{y(-)} respectively.
\subsection{Evaluation Metrics}
In this paper, four metrics: Peak Signal Noise Ratio (PSNR), Structural Similarity (SSIM), Root Mean Square Error (RMSE), and Mean Absolute Error (MAE), are used to measure the watermark embedding and extraction capabilities of the network model.

PSNR is commonly used to evaluate the quality of image reconstruction and is defined by the Mean Square Error (MSE) between two images of size \textit{W}?\textit{H}, \textit{X,} and \textit{Y}. The formula for PSNR is given by:
\begin{equation} \label{14)} 
MSE=\frac{1}{W\times H} \sum _{i=1}^{W}\sum _{j=1}^{H}\left[X_{i,j} -Y_{i,j} \right]  ^{2}  
\end{equation} 
\begin{equation} \label{15)} 
PSNR=10\times \log _{10} \frac{MAX^{2} }{MSE}  
\end{equation} 
In the equation, \textit{X${}_{i,j}$} and \textit{Y${}_{i,j}$} refer to the pixel values of image X and Y at position (\textit{i,j}) respectively. MAX represents the maximum pixel value of an image point, and a higher PSNR value indicates less distortion.

SSIM is another image quality evaluation metric that measures image similarity in terms of brightness, contrast, and structure. It is defined by:
\begin{equation} \label{16)} 
\begin{array}{l} {l(x,y){\rm =}\frac{2\mu x\mu y+C1}{\mu x^{2} +\mu y^{2} +C1} } \\ {{\rm c}(x,y)=\frac{2\sigma x\sigma y+C2}{\sigma x^{2} +\sigma y^{2} {\rm +C}2} } \\ {s(x,y)=\frac{\sigma _{xy} {\rm +}C_{3} }{\sigma _{x} \sigma _{y} {\rm +}C_{3} } } \end{array} 
\end{equation} 
In the equation, \textit{$\mu$${}_{x}$} and \textit{$\sigma$${}_{x}$} are the mean and variance of image \textit{X}, \textit{$\mu$${}_{y}$} and \textit{$\sigma$${}_{y}$} are the mean and variance of image\textit{ Y}, and \textit{$\sigma$${}_{xy}$} is the covariance of X and Y. Constants \textit{C${}_{1}$}, \textit{C${}_{2}$}, and \textit{C${}_{3}$} are used, with \textit{C${}_{1}$=(K${}_{1}$*L)$\wedge$2}, \textit{C${}_{2}$=(K${}_{2}$*L)$\wedge$2}, and \textit{C${}_{3}$=C${}_{2}$/2}. In general, \textit{K${}_{1}$=0.01}, \textit{K${}_{2}$=0.03}, and \textit{L=255}.
\begin{equation} \label{17)} 
SSIM\left(X,Y\right)=l\left(x,y\right)\cdot c\left(x,y\right)\cdot s\left(x,y\right) 
\end{equation} 
SSIM values range from 0 to 1, where a higher value indicates less image distortion.

RMSE indicates the sample standard deviation of the difference between predicted and observed values (called residuals). It is equivalent to the \textit{L${}_{2}$} paradigm and is more sensitive to outliers in the data.
\begin{equation} \label{18)} 
RMSE=\sqrt{MSE}  
\end{equation} 
MAE represents the mean of the absolute errors between predicted and observed values and is equivalent to the \textit{L${}_{1}$} paradigm.
\begin{equation} \label{19)} 
MAE=\frac{1}{W\times H} \sum _{i=1}^{W}\sum _{j=1}^{H}\left|X_{i,j} -Y_{i,j} \right|   
\end{equation} 
\subsection{The imperceptibility of watermarked images}
The imperceptibility of a watermarked image refers to the difficulty for the human eye to detect the presence of a watermark. To achieve blind watermarking, the original image must be visually indistinguishable from the watermarked image. This paper aims to minimize the distortion rate between the original image (i.e., the training image \textit{I}) and the watermarked image \textit{I${}_{W}$}. To evaluate the imperceptibility of the proposed method, four metrics, namely PSNR, SSIM, MAE, and RMSE, are utilized. The experimental results are presented in Table 2.
\begin{table}[b]
    \centering
     \caption{Comparison of embedding rates under different thresholds}
\begin{tabular}{|c|c|c|c|}\hline
\textbf{\textit{Datasets}} & \textbf{\textit{Lego}} & \textbf{\textit{Hotdog}} & \textbf{\textit{Chair}}\\ \hline 
\textbf{Metrics} & \multicolumn{3}{|c|}{\textbf{Training image \textit{I }/ Watermark image\textit{I${}_{W}$}}} \\ \hline 
\textbf{\textit{PSNR}} & \textit{38.226542} & \textit{37.379475} & \textit{37.890828} \\ \hline 
\textbf{\textit{SSIM}} & \textit{0.943185} & \textit{0.918761} & \textit{0.936977} \\ \hline 
\textbf{\textit{MAE}} & \textit{3.168850} & \textit{3.962571} & \textit{3.114965} \\ \hline  \textbf{\textit{RMSE}} & \textit{5.732577} & \textit{6.188527} & \textit{5.956620} \\ \hline  
\end{tabular}

\end{table}

Meanwhile, according to Fig. 6, it is observed that by embedding watermarks in the images of three datasets, namely Lego, Hotdog, and Chair, a comparison between the original training image \textit{I} and the watermarked image \textit{I${}_{W}$} reveals that there is no visible distinction indicating the presence or absence of the embedded watermarking information in the training image. This demonstrates the imperceptibility of the watermark embedded using the method proposed in this paper, resulting in the successful realization of blind watermarking.
\begin{figure}
\centerline{\includegraphics[width=\columnwidth]{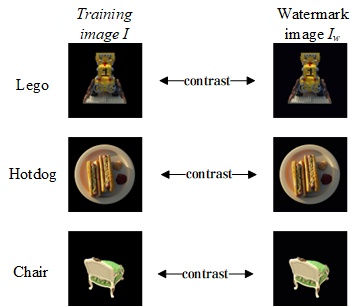}}
\caption{Visual comparison of training images with watermarked images from different datasets.}
\end{figure}

\subsection{Accuracy of watermark extraction}
In this paper, the watermark information, \textit{M${}_{W}$}, is embedded in three datasets, namely Lego, Hotdog, and Chair, using a forward invertible neural network. Then, NeRF is employed for training, and the resulting 3D scenes are rendered to obtain images from different viewpoints. The rendered images are then processed using an image quality enhancement module, after which the recovered watermark, \textit{R${}_{W}$}, is extracted with the aid of an inverse invertible neural network. To evaluate the quality of the extracted watermark information, we used four metrics, namely, PSNR, SSIM, MAE, and RMSE.

\begin{figure}
\centerline{\includegraphics[width=\columnwidth]{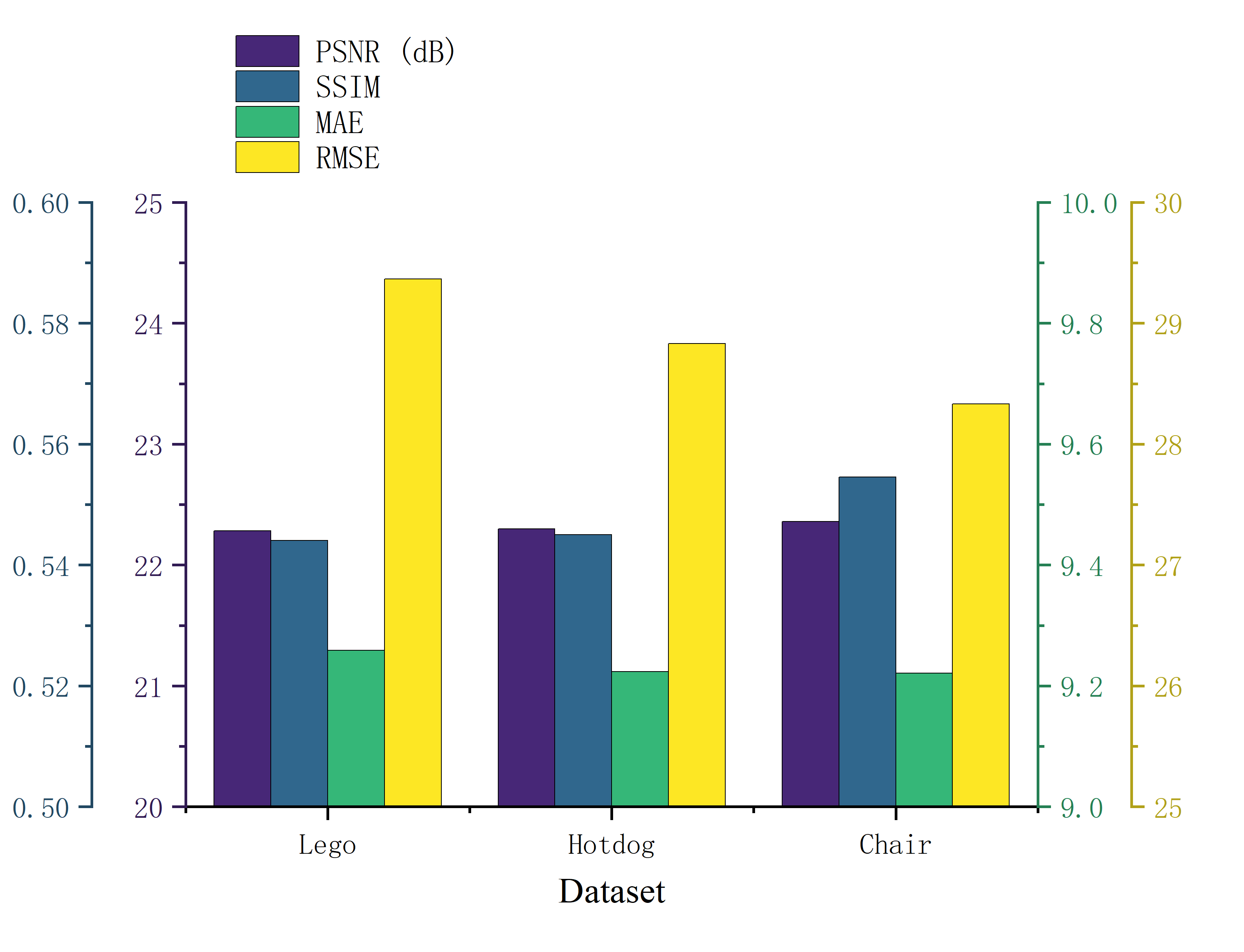}}
\caption{The accuracy of the restored watermark \textit{R${}_{W}$} with the watermark information \textit{M${}_{W}$}.}
\end{figure}
\noindent As depicted in Fig. 7, the average values of each metric for 100 images are as follows: PSNR is greater than 22 dB, SSIM is around 0.55, MAE is around 9.2, and RMSE is approximately 29.
In this study, watermark extraction was conducted on three datasets: Lego, Hotdog, and Chair, for both original training image angles and non-original training image angles. Two parameters, $\Theta$ and $\Phi$, influence the image angle, and this study manipulates $\Theta$ while keeping $\Phi$ constant to control the angle variation. Using the original angles $\Theta$=30, $\Theta$=45, and $\Theta$=60, a view angle offset of +1 is applied to examine whether the watermark information can be successfully extracted when the selected angle differs from the original training angle. The experimental results, depicted in Fig. 8, indicate that the watermark information can be extracted effectively when the selected angle matches the original training image angle. However, when the selected angle is different from the original training image angle (i.e., other angles), the accurate extraction of the watermark information is not achieved.
\subsection{Image Quality Enhancement Module}
Traditional deep learning image robust watermarking techniques, such as HiNet \cite{jing_hinet_2021} and ISN \cite{lu_large-capacity_2021}, are not directly applicable to our task. These techniques rely on reversibility and do not account for the susceptibility of watermarked images to corruption during NeRF rendering. To address this limitation, we propose the incorporation of an image quality enhancement module (IQEM) before the extraction of watermarking operations. By introducing the IQEM, the PSNR of both \textit{M${}_{W}$} and \textit{R${}_{W}$} significantly improves from 5.31dB to 27.23dB, as evidenced in Table 3. The experimental results confirm the efficacy of the IQEM in successfully extracting watermark information.
 
\begin{table}[b]
    \centering
    \caption{Effectiveness of network architecture and design strategies}
    \begin{tabular}{|c|c|c|c|} \hline 
\textbf{\textit{IQEM}} & \textbf{\textit{FDTM}} & \textbf{\textit{L${}_{low-f}$}} & \textbf{\textit{Compare M${}_{W}$${}_{\ }$with R${}_{W}$} (PSNR)} \\ \hline 
$\times$ & $\mathrm{\sqrt{}}$ & $\mathrm{\sqrt{}}$ & 5.31dB \\ \hline 
$\mathrm{\sqrt{}}$ & $\times$ & $\mathrm{\sqrt{}}$ & 12.44dB \\ \hline 
$\mathrm{\sqrt{}}$ & $\mathrm{\sqrt{}}$ & $\times$ & 19.88dB \\ \hline 
$\mathrm{\sqrt{}}$ & $\mathrm{\sqrt{}}$ & $\mathrm{\sqrt{}}$ & 27.23dB \\ \hline 
\end{tabular}

\end{table}

\subsection{Comparison with StegaNeRF}
Modifying the MLP structure to enable watermark embedding is a challenging task. Any direct modifications to the MLP structure can potentially compromise NeRF's rendering ability and hinder its capacity to capture 3D content. Thus, we opt to utilize an invertible neural network watermarking approach for protecting NeRF. Specifically, we embed watermarks onto the 2D images used to train NeRF, and then extract these watermarks from the rendered images to confirm NeRF's copyright. Unlike StegaNeRF, which changes the network structure and may affect NeRF's ability, our proposed method achieves copyright protection through an indirect approach that does not impact the network structure or the rendering ability of NeRF. By training with the same Epoch (50000), our approach yields higher-quality rendered images compared to StegaNeRF, as evidenced in Fig. 8 from a subjective visual perspective.

\begin{figure}
    \centering
    \centerline{\includegraphics[width=\columnwidth]{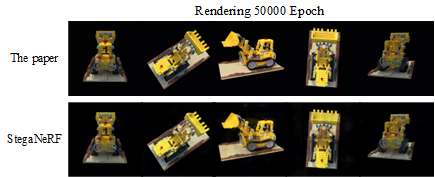}}
    \caption{Visual effects comparison of rendered image quality}
    \label{fig:enter-label}
\end{figure}

The quantitative results of comparing the 13-angle images rendered by two different approaches with the corresponding original training images are presented in Table 3.
\begin{table}[]
    \centering
    \caption{Quantitative comparison of rendered image quality}
   \begin{tabular}{|c|c|c|c|c|} \hline 
\textbf{} & \textbf{PSNR(dB)$\boldsymbol{\mathrm{\uparrow}}$} & \textbf{SSIM$\boldsymbol{\mathrm{\uparrow}}$} & \textbf{MAE$\boldsymbol{\mathrm{\downarrow}}$} & \textbf{RMSE$\boldsymbol{\mathrm{\downarrow}}$} \\ \hline 
\textbf{This paper} & 32.879404 & 0.965815 & 3.021400 & 7.474979 \\ \hline 
\textbf{StegaNeRF} & 29.230964 & 0.907661 & 4.134496 & 10.397353 \\ \hline 
\end{tabular}
\end{table}
The results indicate that our proposed approach outperforms StegaNeRF in all four evaluation metrics. This demonstrates that our approach achieves copyright protection without compromising NeRF's rendering ability.
\section{Conclusion}
In this paper, we propose for the first time a scheme to protect the neural radiation field using invertible neural network watermarking to achieve copyright protection for NeRF. The method employs an invertible neural network to embed and extract watermarks on 2D images, modelling the embedding and extraction of watermarks as forward and reverse processes of the invertible network, while adding an image quality enhancement module in the intermediate process to compensate for the loss of watermark information caused by NeRF rendering process, and to achieve the protection of 3D models represented by neuroradiometric fields. The experimental results show that the scheme in this paper can achieve the embedding and extraction of watermarks, but the extraction quality of watermarks needs to be further improved.

\newpage
\clearpage
\bibliographystyle{IEEEtran}
\bibliography{IEEEabrv,swq}
\end{document}